\documentclass[amsmath,amssymb,preprint]{revtex4-1}
\usepackage{graphicx}
\usepackage{dcolumn}
\usepackage{bm}

\begin{document}

\title{Skyrmion-generated spinmotive forces in inversion broken ferromagnets}
\author{Yuta Yamane\footnote{Corresponding author. \ Email: yuta.yamane@riken.jp}}
\affiliation{ 
RIKEN Center for Emergent Matter Science (CEMS), Wako, Saitama 351-0198, Japan
}

\author{Jun'ichi Ieda}
\affiliation{
Advanced Science Research Center, Japan Atomic Energy Agency, Tokai, Ibaraki 319-1195, Japan
}


\begin{abstract}
We present an analytical study on the spinmotive force (SMF) generated by translational motion of magnetic skyrmion.
A SMF refers to an electrical voltage induced by dynamical magnetic textures, which reflects the spatiotemporal variation of the magnetization.
The dynamics of a skyrmion thus can be detected by a SMF measurement, which may play an important role in future skyrmion-based technologies.
We find the dependence of the SMF on skyrmion structure (e.g., skyrmion or anti-skyrmion, N\'{e}el or Bloch type, and the polarity of the skyrmion core) and Rashba and Dresselhaus spin-orbit couplings (SOCs).
To this end, we derive explicit formulae for the spin-dependent electric fields originating from the two SOCs.
Our findings offer a comprehensive understanding of the phenomenon and an estimation of the electrical voltage signal associated with a moving skyrmion for a given experiment.
\end{abstract}

\keywords{Skyrmion, Spinmotive force, Rashba spin-orbit interaction, Dresselhaus spin-orbit interaction}

\maketitle

\section{Introduction.}
Magnetic skyrmions, since its prediction\cite{Bogdanov,Rossler} and experimental discoveries\cite{Muhlbauer,Yu,Heinze} in magnetic materials with broken inversion symmetry, have been generating increasing attention in the field of spintronics.
They exhibit rich physics stemming from their characteristic, topologically nontrivial structure, and are deemed a promising player for future technologies\cite{Fert,Nagaosa,Finocchio}.
Manipulation of skyrmion dynamics can be achieved by various ways, such as using electrical current\cite{Jonietz,JZhang,Karin}, spin waves\cite{Iwasaki-magnon,Schutte}, temperature gradient\cite{Kong,MochizukiT}, electric field gradient\cite{YHLiu}, and magnetic field gradient\cite{Komineas,SLZhang}.

When a skyrmion is driven into motion, in turn, electrical voltages are induced around the skyrmion\cite{Zhang-skyrm,Ohe2013}, which can be exploited to detect the skyrmion dynamics\cite{Schulz}.
The physics behind such voltage generation has been understood based on the exchange coupling between the conduction electron spin and the local magnetization.
The effect  is known as spinmotive force (SMF)\cite{Berger,Stern,Barnes,Ieda,Hals}.
A SMF reflects the spatiotemporal variation of the magnetization, thus offering a solid way of detecting various dynamical magnetic textures.

In the past decade, the original SMF theory\cite{Korenman,Volovik,Aharonov,Ryu,Yamane2011-jap} has been extended to include the effects of the nonadiabaticity in the electron spin dynamics, i.e., the so-called $\beta$ term\cite{Saslow,Duine,Tserkovnyak}, as well as Rashba spin-orbit coupling (RSOC)\cite{Jalil,Kim,Tatara,Yamane2013}.
Those effects on the skyrmion-generated SMF has been numerically\cite{Shimada} and analytically\cite{Yamane2014,Knoester} studied in earlier work.
There still remain, however, open problems to be addressed.
First of all, the relation between the SMF and the detailed structure of a skyrmion (e.g., skymrion or anti-skyrmion, N\'{e}el or Bloch, and the polarity of the skyrmion core) has not been fully established.
Furthermore, the SMF originating from Dresselhaus spin-orbit coupling (DSOC) has not been discussed in general;
as is well known, a broken bulk (interfacial) inversion symmetry leads to DSOC (RSOC) and the formation of Bloch (N\'{e}el) skyrmions.
The lack of a general estimation of the skyrmion-generated SMF hampers deeper understanding of the phenomenon and its active applications in skyrmion-based technologies.

In this work, we analyze the SMF generated by skyrmion's translational motion, paying a special attention to its dependence on the skyrmion structure and the SOCs.
To discuss systems with broken bulk inversion symmetry as well as interfacial one, we derive a general formula for the SMF originating from DSOC.
We then obtain expressions for the skyrmion-generated SMFs based on a collective-coordinate model, which are supported by a numerical approach.

\section{Spin electric field.}
We assume the Hamiltonian for a conduction electron in the ferromagnet to be given by
\begin{eqnarray}
  H &=& \frac{{\vec p}^2}{2m_{\rm e}} + J {\vec \sigma} \cdot {\vec m} ({\vec r},t)  \nonumber \\ &&
  + \lambda_{\rm D} \left(\sigma_x p_x - \sigma_y p_y\right)
  + \lambda_{\rm R} \left(\sigma_x p_y - \sigma_y p_x\right)  ,
\label{h}
\end{eqnarray}
where $m_{\rm e}$ and ${\vec p}$ are the electron's mass and canonical momentum operator, respectively.
The second term is the exchange coupling, with $J(>0)$ the coupling energy, ${\vec \sigma}$ the vector of Pauli matrices indicating the electron spin operator, and ${\vec m}$ the classical unit vector representing the local magnetization.
The last terms in the second line are the SOCs, with $\lambda_{\rm D}$ and $\lambda_{\rm R}$ are the Dresselhaus and Rashba parameters, respectively.

Equation~(\ref{h}) leads to the spin electric field ${\vec E}^\pm = {\vec E}_0^\pm + {\vec E}_{\rm so}^\pm$, where the upper (lower) sign corresponds to the electrons with majority (minority) spin, and
\begin{eqnarray}
  {\vec E}_0^\pm  &=&   \pm \frac{\hbar}{2e} 
                                 \left( {\vec m} \times \frac{\partial{\vec m}}{\partial t}
                                         + \beta \frac{\partial{\vec m}}{\partial t}
                                 \right) \cdot \nabla {\vec m}  , \label{E0} \\
  {\vec E}_{\rm so}^\pm  &=&  \pm \frac{m_{\rm e}}{e} 
                                         \left[ \left( \frac{\partial {\vec m}}{\partial t} - \beta {\vec m} \times \frac{\partial{\vec m}}{\partial t}\right)_x 
                                                 \left( \lambda_{\rm D} {\vec e}_x + \lambda_{\rm R} {\vec e}_y \right) 
                                                  \right. \nonumber \\ && \left.
                                                 - \left(  \frac{\partial {\vec m}}{\partial t} - \beta {\vec m} \times \frac{\partial{\vec m}}{\partial t}\right)_y
                                                   \left( \lambda_{\rm R} {\vec e}_x + \lambda_{\rm D} {\vec e}_y \right)
                                         \right]  .
\label{Eso}
\end{eqnarray}
Here, $\beta$ is a dimensionless parameter characterizing the non-adiabaticity in the electron spin dynamics, and the SOC parameters are assumed to be time-independent.
See Appendix for a derivation of Eqs.~(\ref{E0}) and (\ref{Eso}).
${\vec E}_0^\pm$ is the SOC-free spin electric field, arising when ${\vec m}$ varies in both time and space.
${\vec E}_{\rm so}^\pm$, on the other hand, is the SOC-induced part.
In addition to RSOC, which has been discussed in earlier work\cite{Jalil,Kim,Tatara,Yamane2013}, here we have also derived the spin electric field originating from DSOC.
The expressions of the contributions from RSOC and DSOC are consistent with what should be expected from the relative symmetry of the two SOCs in Eq.~(\ref{h}).
In contrast to ${\vec E}_0^\pm$, the SOC-induced field ${\vec E}_{\rm so}^\pm$ does not require the spatial variation of ${\vec m}$.

${\vec E}^\pm$ gives rise to the electric current  ${\vec j}_{\rm c} = \sigma_{\rm F}^+ {\vec E}^+ + \sigma_{\rm F}^- {\vec E}^-  = \left( \sigma_{\rm F}^+ - \sigma_{\rm F}^- \right) {\vec E}^+ $, with $\sigma_{\rm F}^{+(-)}$ the electric conductivity for the majority (minority) electrons.
In an open circuit condition, the ordinary electric field ${\vec E}_{\rm ind} = - \nabla\phi - \partial{\vec A}/\partial t$ is induced to keep the total electrical current zero, i.e., ${\vec j}_c +  \left( \sigma_{\rm F}^+ + \sigma_{\rm F}^- \right) {\vec E}_{\rm ind} = 0$, where $\phi $ and ${\vec A}$ are U(1) scalar and vector potentials.
To determine the electric potentials the gauge has to be fixed, and here we adopt the Coulomb gauge, $\nabla\cdot{\vec A} = 0$.
The scalar potential is thus determined by the Poisson equation $ \nabla\cdot P {\vec E}^+ = \nabla^2\phi $, where $P = \left( \sigma_{\rm F}^+ - \sigma_{\rm F}^- \right) / \left( \sigma_{\rm F}^+ + \sigma_{\rm F}^- \right) $.
We define the SMF $V$, induced between given two spatial points ${\vec r}_a$ and ${\vec r}_b $, by the potential difference, i.e., $V = \phi ({\vec r}_b) - \phi ({\vec r}_a)$.

\section{Skyrmion-generated SMF.}
Let us examine the SMF generated by skyrmion motion.
A two-dimensional skyrmion structure, with its center located at $ {\vec r}_{\rm c} = (x_{\rm c} , y_{\rm c})$, is characterized by $\theta({\vec r}-{\vec r}_{\rm c}) = \theta(\rho)$ and $\varphi({\vec r}-{\vec r}_{\rm c}) = c \chi + \varphi_0 $,
where $(\theta,\varphi)$ defines the magnetization direction by ${\vec m} = (\sin\theta\cos\varphi, \sin\theta\sin\varphi, \cos\theta) $;
the two-dimensional cylindrical coordinate $(\rho,\chi)$ is measured from the skyrmion center, i.e., $x-x_{\rm c} = \rho\cos\chi$ and $y-y_{\rm c} = \rho\sin\chi$;
$ c = +1 $ ($-1$) for (anti-)skyrmion;
and $\varphi_0$ is a constant, which has a close relation with the SOCs.
The broken bulk (interfacial) inversion symmetry implies the presence of DSOC (RSOC), which usually leads to a Dzyaloshinskii-Moriya interaction (DMI) that favors a Bloch (N\'eel) skyrmion, i.e., $\varphi_0=\pm\pi/2$ ($0,\pi$).
In the absence of both bulk and interfacial inversion symmetry, a skyrmion adopts a compromise between N\'eel and Bloch structures.
Although the analytical expression for a skyrmion solution is unavailable, it is well approximated by a standard 360$^\circ$ domain wall profile\cite{Romming};
\begin{equation}
 \theta (\rho)  =  4 \arctan\left\{ \exp \left( \frac{q\rho}{w} \right) \right\} + h ,
\label{theta}
\end{equation}
where $q=\pm1$ is the polarity of the skyrmion, $h=-\pi$ ($0$) for $q=+1$ ($-1$), and $w$ defines the skyrmion size.

Throughout this work, we limit ourselves to the steady-motion of a skyrmion with velocity $ {\vec v} = (v,0) = ( dx_{\rm c}/dt, 0 ) $.
Translational motion of a skyrmion can be driven by a variety of methods\cite{Jonietz,JZhang,Karin,Iwasaki-magnon,Schutte,Kong,MochizukiT,YHLiu,Komineas,SLZhang}.
Whatever the driving force is, a dynamical skyrmion induces the spin electric field around it according to Eqs.~(\ref{E0}) and (\ref{Eso}).
In what follows, we derive the SMF in terms of skyrmion velocity, without specifying what drives the skyrmion motion;
it is a trivial task to read the SMF as a function of a particular driving force, once the expression of the skyrmion velocity is known\cite{Note}.

To address analytical expressions for the SMF, here we adopt a ``quasi-one-dimensional'' approximation for electrical voltages, where the longitudinal SMF $V_x$ is estimated by
\begin{equation}
  V_x  =  \frac{1}{L_y} \int_{y_0}^{y_0+L_y} dy \int_{x_0}^{x_0+L_x}  P E^+_x dx  .
\label{Vx}
\end{equation}
Here, $V_x$ is measured between $x=x_0$ and $x_0+L_x$, with two electrodes each extending along the $y$ direction from $y=y_0$ to $y_0+L_y$.
The $y$ integration with the factor $1/L_y$ is to take the spatial-averaging of the electric field below the electrodes.
Equation~(\ref{Vx}) becomes exact when the system is truly one-dimensional.
In the two dimensions, the scheme provides a good approximation when the divergence of the electric field, i.e., the moving skyrmion, is sufficiently away from the electrodes. 
A straightforward calculation leads to $ V_x = V_0^x + V_{\rm so}^x $, where
\begin{eqnarray}
  V_0^x  &=&  -  \frac{v}{L_y} \frac{P\hbar}{2e} \beta \Gamma  , \label{V0_x} \\
  V_{\rm so}^x  &=&  - \frac{v}{L_y} \frac{\pi P m_{\rm e}}{e} 
                                   \left\{ \left(  \lambda_{\rm D} \cos\varphi_0 - \lambda_{\rm R} \sin\varphi_0 \right) I_1
                                   \right. \nonumber \\ && \left.
                                           + \beta \left( \lambda_{\rm D} \sin\varphi_0 + \lambda_{\rm R} \cos\varphi_0 \right) I_2
                                   \right\}  , \label{Vso_x}
\end{eqnarray}
with $ \Gamma = \pi \int \rho d\rho \{ ( \partial\theta / \partial\rho )^2 + ( \sin\theta / \rho )^2 \} $, 
$ I_1 = \int d\rho \{ \rho ( \partial \theta / \partial \rho ) \cos\theta + \sin\theta \} $, and 
$ I_2 = \int d\rho \{ \rho ( \partial \theta / \partial \rho ) + \sin\theta\cos\theta \} $.
With similar concepts, the transverse SMF is given by $V_y = (L'_x)^{-1} \int_{x'_0}^{x'_0+L'_x} dx \int_{y'_0}^{y'_0+L'_y}  P E^+_y dy = V_0^y + V_{\rm so}^y$, where
\begin{eqnarray}
  V_0^y  &=&   - \frac{v}{L'_x}  \frac{P\hbar}{2e} 4\pi N_{\rm S}  , \label{V0_y} \\
  V_{\rm so}^y  &=&  - \frac{v}{L'_x}  \frac{\pi P m_{\rm e}}{e}  
                                   \left\{ \left( \lambda_{\rm R} \cos\varphi_0 - \lambda_{\rm D} \sin\varphi_0 \right) I_1
                                   \right. \nonumber \\ && \left.
                                           + \beta \left( \lambda_{\rm R} \sin\varphi_0 + \lambda_{\rm D} \cos\varphi_0 \right) I_2
                                   \right\}  . \label{Vso_y}
\end{eqnarray}
Here, $V_y$ is measured between $y=y'_0$ and $y'_0+L'_y$, with the two electrodes extending along the $x$ direction from $x=x'_0$ to $x'_0+L'_x$, and $N_{\rm S}$ is defined by
\begin{equation}
  N_{\rm S}  =  \frac{1}{4\pi} \iint dxdy \ {\vec m} \cdot
                             \left( \frac{\partial{\vec m}}{\partial x} \times \frac{\partial{\vec m}}{\partial y} \right) 
                             =  cq .
\label{Ns}
\end{equation}
Using Eq.~(\ref{theta}), the integrals in $\Gamma$, $I_1$, and $I_2$ can be numerically evaluated.
Taking the lower and upper limits of the integrals as $\rho=0$ and $\rho=\infty$, respectively, one finds $\Gamma \simeq 13.5$, $ -4.3\times10^4\times I_1 \simeq | I_2 | $, and $ I_2 \simeq 3 q w$.

Eqs.~(\ref{V0_x})-(\ref{Vso_y}) contain our central results, revealing key features of the skyrmion-generated SMF.
The dc SMFs proportional to the skyrmion velocity $v$ appear both in the $x$ and $y$ directions.
The SOC-free SMFs in Eqs.~(\ref{V0_x}) and (\ref{V0_y}) are independent of $\varphi_0$.
Notice that, in the context of topological Hall effect\cite{Ye,Lee,Neubauer}, $V_0^y$ in Eq.~(\ref{V0_y}) is interpreted as a result of the Lorentz force due to the spin magnetic field, which is related to the spin electric field (\ref{E0}) via the ``Maxwell-Faraday'' equation.
The SOC-induced SMFs in Eqs.~(\ref{Vso_x}) and (\ref{Vso_y}), on the other hand, are proportional to the SOC parameters and trigonal functions of $\varphi_0$.
It should be remarked that $ I_1 $ and $ I_2 $ depend on the Skyrmion size $w$ and so do $V_{\rm so}^x$ and $V_{\rm so}^y$.

\begin{figure*}
  \centering
  \includegraphics[width=13cm, bb=0 0 1022 747]{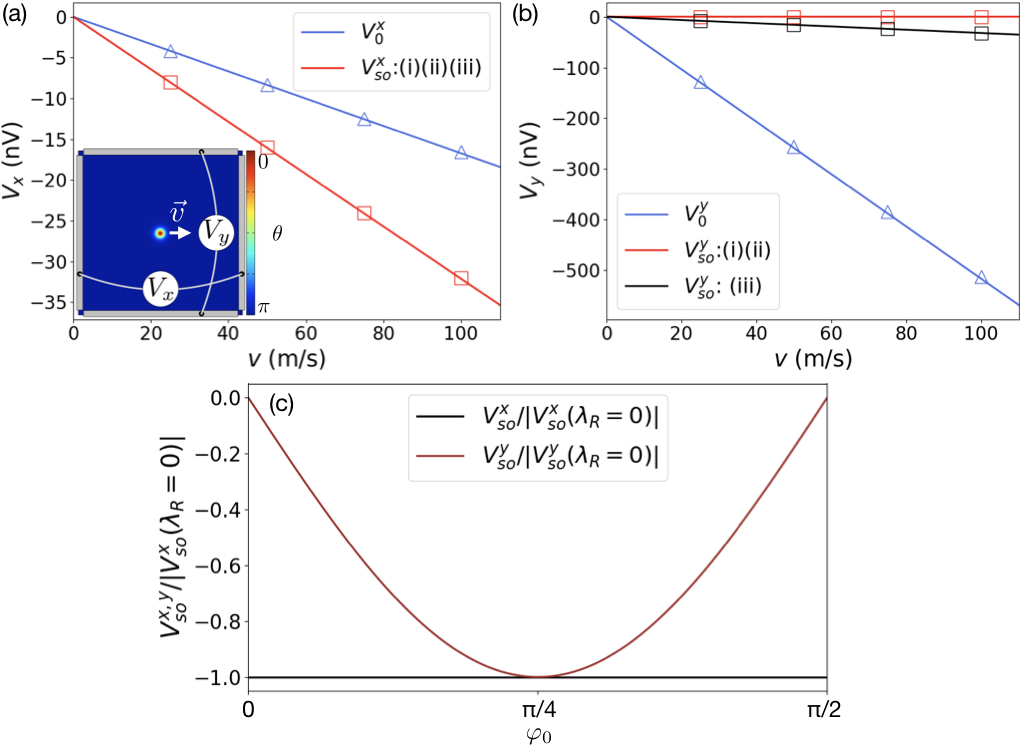}
  \caption{ (a) Longitudinal and (b) transverse SMFs  as functions of skyrmion velocity $v$.
                 The sold lines plot Eqs.~(\ref{V0_x})-(\ref{Vso_y}), while the open symbols are the results obtained by numerically solving the Poisson equations.
                  See the main text for the model parameters used and the combinations of $\varphi_0$ and the SOC parameters corresponding to the three cases (i), (ii), and (iii).
                 The inset of (a) shows the system employed in the numerical simulation.
                 (c) Normalized SOC-induced SMFs, as functions of $\varphi_0$, based on Eqs.~(\ref{Vso_x}) and (\ref{Vso_y}).
                  }
  \label{fig1}
\end{figure*}

In Fig.~\ref{fig1}~(a) and (b), Eqs.~(\ref{V0_x})-(\ref{Vso_y}) are plotted by solid lines as functions of the skyrmion velocity $v$, where three different combinations of $\varphi_0$ and the SOC parameters are examined;
(i) a Bloch skyrmion with $\varphi_0=\pi/2$, $\lambda_{\rm D}= \lambda \equiv 10^{-11}$ eV$\cdot$m$/ \hbar$ and $\lambda_{\rm R} = 0$, (ii) a N\'eel skyrmion with $\varphi_0=0$, $\lambda_{\rm D}=0$ and $\lambda_{\rm R} = \lambda$, and (iii) an intermediate-shape skyrmion with $\varphi_0=\pi/4$ and $\lambda_{\rm D} = \lambda_{\rm R} = \lambda / \sqrt{2}$.
Here we have assumed the proportionality between DMIs and SOCs, leading to the relation $\varphi_0 = \tan^{-1} \left( \lambda_{\rm D} / \lambda_{\rm R} \right)$\cite{Rowland}.
For all the three cases, we assume $P=0.5$, $\beta=0.03$, $m_{\rm e}=9.1\times10^{-31}$ kg, $c=+1$, $q=+1$, $w=10$ nm, and $ L_y = L'_x = 400$ nm.

For the cases (i) and (ii), only the $\lambda_{\rm D}\sin\varphi_0 ( = \lambda )$ and $\lambda_{\rm R} \cos\varphi_0 (=\lambda)$ terms survive, respectively, in Eqs.~(\ref{Vso_x}) and (\ref{Vso_y}).
This leads to the appearance of the same electrical voltage for the two cases, with the vanishingly small $V_{\rm so}^y$ because of the condition $ | I_1 | \ll | I_2 | $.
When the skyrmion takes an intermediate structure, every term in Eqs.~(\ref{Vso_x}) and (\ref{Vso_y}) can contribute.
Especially when $\lambda_{\rm D} = \lambda_{\rm R}$ and $\varphi_0=\pi/4$, as in the case (iii), one finds $V_{\rm so}^x (v) = V_{\rm so}^y (v)$.
In Fig.~\ref{fig1}~(c) , the (normalized) SOC-induced SMFs with a given $v$ are plotted as functions of $\varphi_0$, where we assume $\varphi_0 = \tan^{-1} \left( \lambda_{\rm D} / \lambda_{\rm R} \right) $, fixing the value of $ \lambda = \sqrt{ \lambda_{\rm D}^2 + \lambda_{\rm R}^2 }$.
It is seen that $V_{\rm so}^x$ is independent of $\varphi_0$ while $V_{\rm so}^y$, negligibly small around $\varphi_0\simeq0$ and $\pi/2$, has the peak at $\varphi = \pi /4$ ($\lambda_{\rm D} = \lambda_{\rm R} = \lambda / \sqrt{2}$) where $V_{\rm so}^x = V_{\rm so}^y$.

Now, let us compute SMFs by numerically solving the Poisson equation $ \nabla\cdot P {\vec E}^+ = \nabla^2\phi $, and compare the results with Eqs.~(\ref{V0_x})-(\ref{Vso_y}).
First, to check the analytical results in Fig.~\ref{fig1}~(a) and (b), we consider a square thin film with the side length of $400$ nm [the inset of Fig.~\ref{fig1}~(a)].
We divide the sample into $1\times1$ nm$^2$ unit cells, and assign the magnetization at each unit cell.
The same parameters and skyrmion dynamics as those assumed in the analytical calculation in Fig.~1~(a) and (b) are employed.
The Poisson equation is solved with the boundary condition $\partial \phi / \partial n = 0$, where $n$ is the normal direction to the sample boundary.
The electrodes are located at the sample edges ($x=0$ and $400$ nm for $V_x$, and $y=0$ and $400$ nm for $V_y$) and extends from a corner to another corner [the inset of Fig.~\ref{fig1}~(a)], below which the spatial average of $\phi(x,y)$ is taken. 
Indicated by the open symbols in Fig.~\ref{fig1}~(a) and (b) are the numerical results, which agree well with Eqs.~(\ref{V0_x})-(\ref{Vso_y}).

\begin{figure}
  \centering
  \includegraphics[width=8.5cm, bb=0 0 876 762]{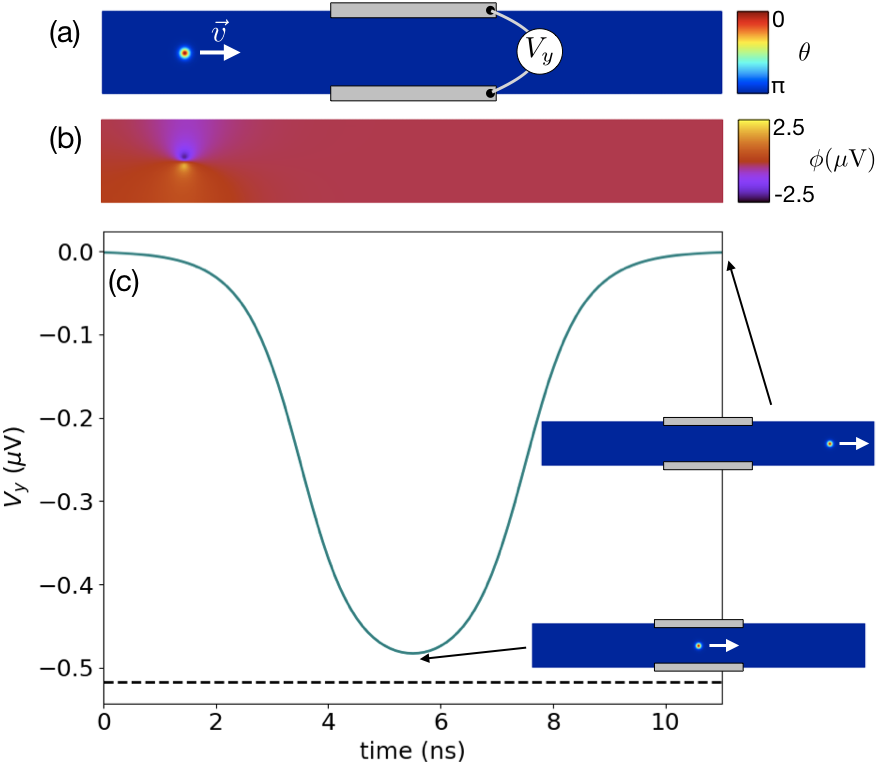}
  \caption{ (a) The initial distribution of the magnetization polar angle $\theta$ in the rectangular sample.
                 The transverse SMF $V_y$ is measured as indicated, during the skyrmion propagation to the right.
                 (b) Electric potential distribution numerically obtained from the magnetization profile in (a).
                 (c) Temporal evolution of $V_y$.
                 The insets indicate the skyrmion positions at $t=5.5$ ns and $11$ ns.
                 See the main text for the model parameters used.
            }
  \label{fig2}
\end{figure}

Next, we consider a rectangular sample with the dimensions of $1500\times200$ nm$^2$, containing a skyrmion that is initially located at $(x_{\rm c},y_{\rm c})=(200\ {\rm nm},100\ {\rm nm})$ and exhibits the steady motion [Fig.~\ref{fig2}~(a)].
While the Magnus force generally hinders a skyrmion to be propagated in the intended direction\cite{Jiang,Litzius}, we here assume, for simplicity, the guided motion of skyrmion along the $x$ direction, by tuning the spatial dependence of the sample thickness\cite{Purnama}.
As the skyrmion travels from $x_{\rm c}=200$ nm to $1300$ nm with $v=100$ m/s, the transverse SMF $V_y$ is monitored with the electrodes placed at $y=0$ and $200$ nm, each extending from $x=550$ nm to $950$ nm.
For the material parameters and skyrmion profile, those used for the N\'eel skyrmon in Fig.~\ref{fig1} are employed.
Since the major electric-potential drop takes place around the skyrmion [Fig.~\ref{fig2}~(b)], $V_y$ is finite only when the skyrmion is moving around the area sandwiched by the electrodes.
In Fig.~\ref{fig2}~(c), the temporal evolution of $V_y$ is recorded by the solid curve, showing the peak when the skyrmion reaches the middle of the sample.
The dotted line in Fig.~\ref{fig2}~(c) indicates the analytical result [the sum of Eqs.~(\ref{V0_y}) and (\ref{Vso_y})] with $L'_x=400$ nm.
Remark that Eqs.~(\ref{V0_x}) to (\ref{Vso_y}) have been derived implicitly assuming that the electrodes are placed across the moving skyrmion.
The quantitative difference between the analytical prediction and the peak value of the numerical result, about six percent, is attributed to the fact that, in the simulation, $\phi$ can decay beyond the region between the electrodes, i.e., $x<550$ nm and $x>950$ nm.

When the skyrmion motion is deflected from the direction of the long side of the rectangular sample, the ``transverse'' voltage  will be given by a linear combination of Eqs.~(\ref{V0_x}) to (\ref{Vso_y});
notice that the ``x'' and ``y'' directions in Eqs.~(\ref{V0_x}) to (\ref{Vso_y}) are defined with respect to the skyrmion motion.
This may hamper an accurate estimation of the SMF when the Magnus force on the skyrmion makes it difficult to precisely predict its motion\cite{Jiang,Litzius}.
In such cases, one can compare Eqs.~(\ref{V0_x}) to (\ref{Vso_y}) with the experimentally-observed longitudinal and transverse voltages, to track the direction of the skyrmion motion as well as to draw qualitative information on the skyrmion structure and the SOC parameters.

In the analyses in Figs.~\ref{fig1} and \ref{fig2}, we employed $c=+1$ for definiteness.
An anti-skyrmion is characterized by $c=-1$, on the other hand, leading to the reverse of the sign of $V_0^y$ in Eq.~(\ref{V0_y}) through its dependence on $N_{\rm S}$.
As $V_0^y$ is the dominant contribution to the transverse voltage over $V_{\rm so}^y$, the SMF offers a way to electrically distinguish a skyrmion and an anti-skyrmion.

\section{Discussions and conclusions.}
To address the analytical expressions for the SMF, we have assumed the steady-motion of skyrmion sufficiently away from the sample boundaries.
The investigation on the influence of skyrmion-shape distortion, transient dynamics, the Magnus force, etc. must be interesting, and it will stimulate further systematic study based on micromagnetic simulations in future.
It is also known that in general a dynamical magnetic texture leads to feedback effects on the magnetization dynamics\cite{Kim,ZhangZhang,Utkan}, which have been neglected in our steady-motion model.
The assumptions made in deriving Eqs.~(\ref{V0_x}) to (\ref{Vso_y}) should be met when the driving force is sufficiently moderate, so that the collective-coordinate approach for the skyrmion dynamics is appropriate.

With the sets of parameters used in the present study, $V_0^x$ is comparable to $V_{\rm so}^x$, while $V_0^y$ is dominant over $V_{\rm so}^y$ [Fig.~\ref{fig1}~(a) and (b)].
A quite large SOC parameter has been reported, however, for Pt/Co/AlOx\cite{Miron}, where $\lambda_{\rm R} = 10^{-10} $ eV$\cdot$m$/\hbar$;
this is ten times larger than the value we assumed in Fig.~\ref{fig1}~(a) and (b).
We therefore expect that for certain systems the SOC-induced SMF can provide the most important contribution to the observed electrical voltage.
Larger SMFs may be realized also by exploiting a smaller sample, since they are inversely proportional to the sample dimensions [the factor $1/L_y$ in Eqs.~(\ref{V0_x}) and (\ref{Vso_x}), and the factor $1/L'_x$ in Eqs.~(\ref{V0_y}) and (\ref{Vso_y})].

We have assumed the relation $\varphi_0 = \tan^{-1}\left( \lambda_{\rm R} / \lambda_{\rm D} \right) $ between the skyrmion structure and the SOCs, i.e., a DSOC (RSOC) leads to a Bloch (N\'eel) skyrmion.
Recently, however, the possibility of coexistence of RSOC and Bloch skyrmion has been predicted\cite{Motome}.
In this case, according to Eqs.~(\ref{Vso_x}) and (\ref{Vso_y}), $V_{\rm so}^x(v)$ is negligibly small compared to $V_{\rm so}^y (v)$, on the contrary to the case (ii) in Fig.~\ref{fig1}~(a) and (b). 
Equations~(\ref{V0_x}) to (\ref{Vso_y}) can be just applied to arbitrary combinations of SOC and skyrmion structure.

So far we have considered the SMFs generated by a single skyrmion.
When multiple skyrmions are involved\cite{Ohe2013,Shimada}, the net electric potential distribution is given by the superposition of all individual electric potentials produced by each skyrmion.
Let us assume that $N$ identical skyrmions move coherently, keeping sufficiently large distance from each other so that the electric potential distribution around each skyrmion is little affected by the presence of the other skyrmions. 
In this case, a rough estimation for the longitudinal and transverse SMFs, $V_x^N$ and $V_y^N$, respectively, are given by $V_x^N = N V_x$ and $V_y^N = N V_y$,\cite{Yamane2014} where $V_x$ and $V_y$ are the SMFs induced by a single skyrmion. 
Numerical simulations of the Poisson equation (not shown) support this prediction.

In conclusion, we have studied the SMF generated by translational motion of magnetic skyrmion.
We have derived analytical expressions for the SMFs in the presence of RSOC and DSOC, which enable to give an estimation of the electrical voltages for a given skyrmion system.
Our results offer a fundamental understanding of the phenomenon, revealing the dependence of the SMF on the SOC parameters and skyrmion's structure;
e.g., skyrmion or anti-skyrmion, N\'eel or Bloch, and the polarity of the skyrmion core.
Our results, enabling to estimate the SMF for a given skyrmion profile, paves a path to the implementation of the phenomenon in future spintronics applications.

\section*{Acknowledgments}
The authors appreciate K. Yamamoto and B. F. McKeever for valuable comments on the manuscript.
This research was supported by Research Fellowship for Young Scientists (No.~17J03368) from Japan Society for the Promotion of Science (JSPS), and Grant-in-Aid for Scientific Research (C) (No.~16K05424) from JSPS.

\section*{Appendix: Derivation of spin electric field}
To derive the spin electric field in Eq.~(\ref{E0}) and (\ref{Eso}), we consider the Heisenberg equation of motion\cite{Yamane2011-jap,Yamane2013}.
The velocity operator ${\vec v}$ of the electron is given by ${\vec v} = \left( 1 / i \hbar \right) \left[{\vec r}, H \right] = {\vec p} / m_{\rm e} +  \lambda_{\rm D} {\cal M} \left[ {\vec \sigma} \right] - \lambda_{\rm R} {\vec \sigma} \times {\vec e}_z $, where $ {\cal M} \left[ {\vec \sigma} \right] = ( \sigma_x, - \sigma_y, 0 )$.
The force operator ${\vec F}$ is defined as ${\vec F} = \left( 1 / i \hbar \right) \left[ m_{\rm e} {\vec v}, H \right] + \partial \left( m_{\rm e}{\vec v} \right) / \partial t $, the explicit expression of which is given by
\begin{eqnarray}
  {\vec F}  &=&  - J  \nabla \left( {\vec \sigma} \cdot {\vec m} \right)
                         + \frac{m_{\rm e}}{i\hbar} \left[ \lambda_{\rm D} {\cal M} \left[ {\vec\sigma} \right] - \lambda_{\rm R} {\vec\sigma}\times{\vec e}_z, J {\vec \sigma}\cdot{\vec m} \right]  \nonumber \\ &&
                         + m_{\rm e} \left( \frac{\partial\lambda_{\rm D}}{\partial t} {\cal M} \left[ {\vec \sigma} \right] - \frac{\partial\lambda_{\rm R}}{\partial t} {\vec \sigma} \times {\vec e}_z \right)  .
\label{f}
\end{eqnarray}
The first term is interpreted as the spatial gradient of the ``potential energy'' $J{\vec \sigma}\cdot{\vec m}$.
The second term originates from the non-commutating nature of the SOCs and the exchange interaction.
The third term is induced by the temporal variation of the ``vector potentials'' due to the SOCs.
To calculate the electrical voltage, we have to determine the expectation value of ${\vec F}$.
In Eq.~(\ref{f}) and hereafter, we discard ${\vec v}$-dependent terms, anticipating that we will consider open circuit conditions where the expectation value of ${\vec v}$ vanishes. 
Notice that, however, when the Skyrmion motion is driven by in-plane electrical current, the circuit is closed.
 We can thus no longer ignore the ${\vec v}$-dependent terms in ${\vec F}$, which would give rise to extra contributions to the force.
For the particular case of current-driven Skyrmion motion a more careful analysis will be therefore required, which will be discussed separately.

Let us determine the (normalized) electron spin ${\vec s}_\pm = \langle {\vec \sigma} \rangle_\pm$, where $\langle ... \rangle_{+(-)}$ stands for the expectation value for the electron at Fermi surface with the majority (minority) spin.
We here assume the condition $ J / \hbar \gg \lambda_{{\rm D},{\rm R}} | {\vec k}_{\rm F} | $, with ${\vec k}_{\rm F}$ the Fermi wave vector, so that the electron spin dynamics is dictated by the exchange coupling.
If the spatiotemporal variation of the magnetization is slow enough for the electron spin to adiabatically follow, ${\vec s}_\pm$ may thus be decomposed as 
\begin{equation}
  {\vec s}_\pm  =  \mp {\vec m} + \delta{\vec s}_\pm  ,
\label{s}
\end{equation}
where $\delta{\vec s}_\pm$ ($| \delta{\vec s}_\pm | \ll 1$) represents the small deviation of the electron spin from $\mp {\vec m}$.
We assume the electron spin to obey the equation of motion\cite{ZhangLi} 
\begin{equation}
  \frac{\partial{\vec s}_\pm}{\partial t}  =  - \frac{1}{\tau_{\rm ex}} {\vec s}_\pm \times {\vec m} 
                                             - \frac{1}{\tau_{\rm sf}} \delta{\vec s}_\pm  ,
\label{eom}
\end{equation}
where the first term in the rhs represents the Larmor precession around the magnetization, with $\tau_{\rm ex} = \hbar/2J$, while the second term phenomenologically describes the damping motion towards $\mp{\vec m}$, with $\tau_{\rm sf}$ being the spin relaxation time.
Substituting Eq.~(\ref{s}) into (\ref{eom}), we obtain $\delta{\vec s}_\pm$ in terms of ${\vec m}$ as\cite{ZhangLi} 
\begin{equation}
  \delta{\vec s}_\pm  =  \pm \frac{\tau_{\rm ex}}{1+\beta^2}
                                     \left( {\vec m} \times \frac{\partial{\vec m}}{\partial t} + \beta \frac{\partial{\vec m}}{\partial t}  \right)  , 
\label{ds}
\end{equation}
where $\beta = \tau_{\rm ex} / \tau_{\rm sf}$.
The adiabatic condition for the electron spin dynamics indicates $\tau_{\rm ex} \ll \tau_{\rm sf}$, i.e., $\beta \ll 1$.

The expectation value of ${\vec F}$ in Eq.~(\ref{f}) is given, using Eq.~(\ref{s}) with (\ref{ds}), as
\begin{equation}
    \langle {\vec F} \rangle_\pm = \pm \left\{ - e \left( {\vec E}_0 + {\vec E}_{\rm so} \right) \right\}  ,
\end{equation}
where the SOC-free and SOC-induced spin electric fields, ${\vec E}_0$ and ${\vec E}_{\rm so}$, respectively, are defined by Eqs.~(\ref{E0}) and (\ref{Eso}).
Here, the SOC parameters are assumed to be time-independent.
While our treatment of the electron spin relaxation in Eq.~(\ref{eom}) is phenomenological, the resulting $\beta$ terms in Eqs.~(\ref{E0}) and (\ref{Eso}) are consistent with the microscopic derivations in the previous work.\cite{Saslow,Duine,Tatara}



\begin{thebibliography}{99}
\bibitem{Note}  When the Skyrmion motion is driven by in-plane electrical current, there can appear additional forces that depend on the electron velocity (see Appendix for a more detailed discussion). For the particular case of current-driven Skyrmion motion a more careful analysis will be therefore required, which will be discussed separately.
\end{thebibliography}
\end{document}